\documentclass[11pt]{article}
\usepackage{geometry}                
\usepackage[parfill]{parskip}    
\usepackage{graphicx}
\usepackage{amssymb}
\usepackage{epstopdf}

\long\def\symbolfootnote[#1]#2{\begingroup%
\def\thefootnote{\fnsymbol{footnote}}\footnote[#1]{#2}\endgroup}

\title{Dark Matter Capture in the First Stars: a Power Source 
and Limit on Stellar Mass}
\author{
\mbox{Katherine Freese$^{1,2}$\footnote{Email:ktfreese@umich.edu},}
\mbox{Douglas Spolyar$^3$\footnote{Email:dspolyar@physics.ucsc.edu},}
and
\mbox{Anthony Aguirre$^3$\footnote{Email:aguirre@scipp.ucsc.edu}}
}

\begin{document}
\maketitle

\begin{center}
$^{1}$Michigan Center for Theoretical Physics, Physics Dept.,
Univ. of Michigan, Ann Arbor, MI 48109\\ 
$^{2}$Perimeter Institute for Theoretical Physics, 31 Caroline St. N. Waterloo
Ontario, Canada N2L 2Y5\\ 
$^{3}$Physics Dept., University of California, Santa Cruz, CA 95060
\end{center}

\begin{abstract}
\noindent
The annihilation of weakly interacting massive particles can provide an important heat
source for the first (Pop. III) stars, potentially leading to a new phase of
stellar evolution known as a ``Dark Star".  When dark matter (DM) capture
via scattering off of baryons is included, the luminosity from DM
annihilation may dominate over the luminosity due to fusion, depending
on the DM density and scattering cross-section. The influx
of DM due to capture may thus prolong the
lifetime of the Dark Stars.  Comparison of DM luminosity with
the Eddington luminosity for the star may constrain the stellar mass
of zero metallicity stars; in this case
DM will uniquely determine the mass of the first stars.
Alternatively, if sufficiently massive Pop. III stars are found, they
might be used to bound dark matter properties.

\end{abstract}

\section{Introduction}

The first stars in the Universe mark the end of the cosmic dark ages,
reionize the Universe, and provide the enriched gas required for later
stellar generations.  They may also be important as precursors to
black holes that coalesce and power bright early quasars.  The first
stars are thought to form inside dark matter halos of mass $ 10^5
M_\odot$--$ 10^6 M_\odot$ at redshifts $z=10-50$
\cite{Yoshida03,Reed:2005fb}.  These halos consist of 85\% dark matter
and 15\% baryons in the form of metal-free hydrogen and helium gas.
Theoretical calculations indicate that the baryonic matter cools and
collapses via molecular hydrogen cooling
\cite{Peebles:1968nf,matsuda,Hollenbach} into a single small protostar
\cite{omukai} at the center of the halo (for reviews see
e.g. \cite{Ripamonti:2005ri,Barkana:2000fd,Bromm:2003vv}).

In this paper we continue our previous work \cite{sfg} examining the
effect of dark matter (DM)the particles on the first stars.  We focus
on Weakly Interacting Massive Particles (WIMPs), which are the
favorite dark matter candidate of many physicists because they
automatically provide approximately the right amount of dark matter,
i.e.  24\% of the current energy density of the Universe.  Many WIMP
candidates are their own antiparticles, in which case they can
annihilate with themselves in the early universe with weak interaction
cross sections, leaving behind this relic density.  Probably the best
example of a WIMP is the neutralino, which in many models is the
lightest supersymmetric particle. \footnote{If the lightest
supersymmetric particle were the axino or gravitino, it would not have weak interaction annihilations and would not produce the
phenomena described in this paper.}  The neutralino, the supersymmetric
partner of the W, Z, and Higgs bosons, has the required weak
interaction cross section and mass $\sim$ GeV - TeV to give the
correct amount of dark matter and would play an important role in the
first stars.  For reviews of SUSY and other dark matter candidates see
\cite{jkg,bhs,hp}.  

This same annihilation process is the basis of the work we consider
here.  WIMP self-annihilation is relevant wherever the WIMP density is
sufficiently high.  Such regimes include the early Universe, in
galactic halos today \cite{ellis,gs}, in the Sun \cite{sos} and Earth
\cite{freese,ksw}, and in the first stars. As our canonical values, we
will use the standard value for the annihilation cross section 
\begin{equation}
\label{eq:anncrossection}
\langle \sigma v \rangle_{\rm ann} = 3 \times 10^{-26} {\rm cm}^3/{\rm
s}, 
\end{equation}
as this gives the right WIMP relic density today,
as well as take $m_\chi
= 100\,{\rm GeV}$ for our canonical value of the WIMP particle mass, 
but will also consider a
broader range of WIMP masses (1 GeV--10 TeV) and
cross-section.  

The interaction strengths and masses of the
  neutralino depend on a large number of model parameters. In the
  minimal supergravity model, experimental and observational bounds
  restrict $m_\chi$ to 50 GeV--2 TeV, while $\langle \sigma v \rangle$
  lies within an order of magnitude of $3 \times 10^{-26}\, {\rm
    cm^3/sec}$ (except at the
 low end of the mass range where it could
  be several orders of magnitude smaller) \cite{msugra,dsusy}. 
Nonthermal particles can
  have annihilation cross-sections that are many orders of magnitude
  larger (e.g.~\cite{moroi}) and would have even more drastic effects.
  Given the present state of the field there are many types of DM
  candidates which could apply~\cite{fargion}; the effects we find
apply equally well to other WIMP candidates with comparable cross
sections for self-annihilation and scattering off of nucleons.

In this paper we consider the effects of WIMP annihilation on the
first stars.  In a previous paper (hereafter, Paper I) \cite{sfg} two
of us (together with P. Gondolo) considered the effects of dark matter
annihilation on the formation of the first stars.  We found that a
crucial transition takes place when the gas number density of the collapsing
protostar exceeds a critical value ($10^{13}\, {\rm cm}^{-3}$ for a
100 GeV WIMP mass): at this point WIMP annihilation heating dominates
over all cooling mechanisms and prevents the further collapse of the
star. We suggested that the very first stellar objects might be ``dark
stars,'' a new phase of stellar evolution in which the DM -- while
only supplying 1\% of the mass density -- provides the power source
for the star through DM annihilation.  We are as yet uncertain of the
lifetime of these theoretical objects. Once the DM contained inside
the star runs out, the star could contract and heat up to the point
where fusion becomes possible.  However, as the star reaches nuclear
density, there is an additional mechanism for repopulating the DM
inside the star: capture of more DM from the ambient medium.  This new
source of DM can extend the lifetime of the Dark Star.  Indeed the
capture process can continue as long as there is enough ambient DM
passing through the star, causing DM annihilation inside the star to
continue, possibly even to today.

Dark Stars, powered by DM annihilation, require three key ingredients,
as shown in Paper I: 1) high dark matter density, 2) annihilation
products remain trapped in the star, and 3) DM heating dominates over the
other sources of heating.  The first stars exist at the
right place and at the right time to have the best chance of achieving
the first criterion: they exist in the high density centers of dark
matter haloes, and they form at high redshifts (density scales as
$(1+z)^3$).  These densities are still not enough: Paper I showed that
the DM density must be driven up still further in order for the DM
heating to become important: as the baryons condense to form stars,
they come to dominate the potential well and pull the DM in with them
and drive up the density.  The resultant DM densities were computed in
Paper I by adiabatic contraction.  Once the DM due to this effect runs
out, the Dark Star phase might end.  On the other hand, there is
another effect that can repopulate the DM inside the stars and allow
the Dark Star phase to continue: capture of more DM from the ambient
medium.  To get a significant amount of captured material into the
star requires the enhanced densities due to adiabatic contraction. We
will estimate the required ambient DM density for capture to be
important, and find that it is somewhat lower than is required in the
earlier protostellar phase.

In this paper, we consider the effects of DM annihilation on early
zero-metallicity (Population III) stars, once they do have fusion
inside their cores.  These stars live inside a reservoir of WIMPs; as
the WIMPs move through the stars, some of the WIMPs are captured by
the stars.  The captured DM sinks to the center of the stars, where
the DM can annihilate very efficiently.  This has the effect of
dramatically increasing the annihilation rate inside of a star,
compared to DM annihilation without scattering.  The annihilation
provides a heat source for the stars, and we compare this DM
annihilation luminosity to the fusion luminosity of the stars, as well
as to the Eddington luminosity.  Of course, the presence of DM (with
high densities due to adiabatic contraction or capture) would already
become important during (and seriously affect) the formation and
earlier stages of these stars.  Here it is simply our intent to show
that under some circumstances DM capture makes the importance of DM
heating in Pop. III stars unavoidable. Thus, even if DM annihilation
fails to stop a Pop. III star from forming (as was considered in
\cite{sfg}), a Dark Star powered by DM annihilation may exist at the
high nuclear densities where fusion can take place.  Again, we wish to
note that the DM supplies less than a percent of the mass of the Dark
Star and yet may be responsible for its luminosity.

The two key uncertainties in this work are: (i) the scattering cross
section must be at (or near) the experimental limit, and (ii) the
ambient DM density must be high enough for capture to take place.
Whereas the annihilation cross section is fixed to be close to
Eq.(\ref{eq:anncrossection}), on theoretical grounds the scattering cross section
can vary across many orders of magnitude.  As discussed later, it is,
however, constrained by experimental bounds.  For scattering to matter
in the first stars, the scattering cross section must be within two
orders of magnitude of the current experimental limits.  Such a cross
section should be experimentally accessible in the next round of DM
detection experiments.  The second criterion is likely to be true for
a while after the star is created, but it is not clear for how long.
Once the halo containing the Dark Star merges with other objects, it
is not clear how long the central DM in the halo remains undisturbed,
and it is not clear how long the Dark Star remains at this central
point.  In principle the capture could continue indefinitely so that a
dark star could still exist today, but this is very unclear.

Previous work on DM annihilation powering stars has also been done in
the context of high DM densities near the supermassive black holes in
galactic centers, e.g. WIMP burners \cite{moskalenko} and more
generally \cite{edsjo,bertone2007}.

Just as we were preparing to submit our paper, a very similar work was
submitted by~\cite{iocco}.  We have carried the analysis
further. We agree with~\cite{iocco} in the conclusion that DM
annihilation may dominate over fusion, and we illustrate the DM
densities required for this conclusion.  In addition, we go one step
further and discuss the possibility that the DM power source may
exceed the Eddington luminosity and prevent the first stars from
growing beyond a limited mass. This would effect the IR background,
the re-ionization of the universe, the number of supernova, and
potentially the nature of supernova of the first stars; this will be
addressed in a separate publication~\cite{woosley2}.

We begin by discussing the equilibrium WIMP abundance in the first
stars, by computing the number of WIMPs captured by the first stars,
and equating this with the annihilation rate of WIMPs in the first
stars.  A discussion of adiabatic contraction, which may drive up the
DM density near the baryons, follows.  Then we compute the DM
annihilation luminosity and compare with fusion luminosity. Finally,
we compare the DM luminosity with the Eddington luminosity and find a
maximum stellar mass as a function of DM density.

 \section{{WIMP Abundance}}

As WIMPs travel through a star, they can scatter off of the nuclei in
the star with the scattering cross section $\sigma_{c}$.  Although
most of the WIMPs travel right through the star, some of them lose
enough energy to be captured. We call the capture rate $C (s^{-1})$.
The WIMPs then sink to the center of the star, where they can
annihilate with one another with the annihilation rate $\Gamma_A
(s^{-1})$. 
 This process was previously noticed as important for the
Sun by \cite{sos} and in the Earth by \cite{freese,ksw}. DM could 
 be detected through neutrinos produced from the DM annihilation products 
in the Sun or Earth with
experiments such as SuperKamiokande \cite{superk} and 
AMANDA (which did not find a signal and placed bounds)
and ICECUBE (which is starting to take data)\cite{Hubert:2007zzc}.  
Other indirect searches such as GLAST and PAMELA could
detect the gamma-ray and positron annihilation products respectively
from DM annihilating in the Milky Way halo (the gamma-rays and positrons
from DM annihilating in the Sun or Earth would be trapped inside the
objects and would not make it to our detectors).

The number of WIMPs $N$ in the star is then determined by a
competition between capture and annihilation via the differential
equation, 
\begin{equation}
\label{eq:anntwo}
\dot N = C - 2\Gamma_A \equiv C- C_A N^2 , 
\end{equation} 
where $\Gamma_A$
is the annihilation rate (and the factor of two appears because two
particles are annihilated in each event), and 
\begin{equation}
\label{eq:independent}
C_A=2\Gamma_A/N^2 
\end{equation}
is defined as an
$N-$independent annihilation coefficient.  Solving this equation,
we find
 $\Gamma_A = {1 \over 2} C {\rm tanh}^2(t/\tau)$, where
\begin{equation}
\label{eq:equiltime}
\tau = (C C_A)^{-1/2}
\end{equation}
is the equilibration timescale.  Equilibrium corresponds to a balance
between the capture and annihilation rates, i.e.
\begin{equation}
\label{eq:inequil}
\Gamma_A = {1 \over 2} C. 
\end{equation}
As we will show, for the case of Pop III stars, 
equilibrium is quickly reached ($t\gg\tau$) 
and we may use Eq.(\ref{eq:inequil}).

\subsection{Annihilation Rate}

The annihilation rate 
\begin{equation}
\Gamma_A = \int d^3r\, n_\chi(r)^2 (\sigma v)_{\rm ann}
\end{equation}
where $n_\chi(r)$
is the density of captured DM at a point $r$ inside of the star.
In this section we will show that the WIMPs quickly
thermalize with the core of the star,
so that we can treat the DM density distribution as isothermal,
and then we will compute the annihilation rate.
Our work closely follows the approach previously given by
\cite{griestseckel}.

First let us examine the WIMP thermalization timescale inside the star.
The amount of energy $\Delta E$
lost by a WIMP in a scattering event with a nucleus (proton mass
$M_p$) in the star ranges
from $0 \leq {\Delta E \over E} \leq {m_\chi M_p \over [(m_\chi + M_p)/2]^2}$.
Assuming a flat distribution and taking $M_p \ll m_\chi$, the average
energy loss is $2 M_p / m_\chi$.  Roughly, thermalization requires
$\Delta E/E \sim 1$; i.e. there must be $m_\chi / 2 M_p$ scatters.
Thus, for $m_\chi \gg 1$ GeV, 
the timescale for thermalization can be estimated as
\begin{equation} 
\label{time_th}
\tau_{th}\approx \frac{1}{\sigma_c v_{\rm esc} n_H}\frac{M_\chi}{2M_H}
\end{equation}
where $v_{\rm esc}$ is the escape velocity of a DM particle from the surface of
the star, and $n_H$ is the average density of the star.  For a hundred
GeV WIMP, $\sigma_c = 10^{-39}\, {\rm cm}^{-2}$, and $n_H = 10^{24}\,{\rm cm^{-3}}$, we
find that the thermalization time scale to be very short, roughly three
months.

Thus we can use an isothermal distribution for the DM,
\begin{equation}
\label{eq:DMrho}
n(r)_\chi=n_ce^{-m_\chi\phi/kT}
\end{equation}
where $n_c$ is the central number density of DM and
$T$ is the central temperature of the star,
\begin{equation}
\label{eq:phi}
\phi(r)=\int_0^r \frac{GM(r)}{r}dr.
\end{equation}
is the gravitational potential at radius $r$ with respect
to the center, and $M(r)$ is
the mass interior to $r$. One can define effective volumes
\begin{equation}
\label{eq:effVol}
V_j = 4\pi\int_0^{R_*}r^2 e^{-j m_\chi \phi/T}{\rm d}r.
\end{equation}
Upon integration this gives
\begin{equation}
V_j = [3 m_{\rm pl}^2 T/(2jm_{\chi} \rho_c)]^{3/2},
\end{equation}
where $m_{\rm pl}$ is the Planck mass, and $\rho_c$ is the core mass density
of the star.  The name ``effective volume'' is suggestive since we have
$N=n_oV_1$ and also the total annihilation rate is given as
$\Gamma=\langle \sigma v \rangle_{\rm ann}n_o^2V_2$.  

One can then solve Eq.(\ref{eq:anntwo}) to find the
N-independent annihilation coefficient 
defined in Eq.(\ref{eq:independent}),
\begin{equation}
C_A = \langle \sigma v \rangle_{\rm ann} {V_2 \over V_1^2} .
\end{equation}

\begin{table*}
\begin{tabular}{lccccc}
  $M_\star(M_\odot)$ & T$(K)$& $\rho ({\rm g/cm^3})$ & $V_2/V_1^2({\rm cm^{-3}})$&$C_A(s^{-1})$ \\ \hline Sun &
  -&- &$1.72\times10^{-28}$&$5.16\times10^{-54}$ \\ 10& $9.55\times10^{7}$
  &$225.8$&$1.77\times10^{-29}$&$5.31\times10^{-55}$ \\ 50&$1.13\times10^8$	
  &$48.63$&$1.38\times10^{-30}$&$4.14\times10^{-56}$ \\ 100& $1.18\times10^8$
  &$31.88$&$6.86\times10^{-31}$&$2.06\times10^{-56}$ \\ 250&$1.23\times10^8$& $19.72$&
  $3.11\times10^{-31}$&$9.33\times10^{-57}$\\
\end{tabular}
\caption{ \label{tab:PoPIII} Central temperature $T(K)$ and central
  baryon density $\rho(g/cm^3)$ for various masses of metal free stars
  half way through hydrogen burning \cite{woosley}.  
  The effective volume $V_2/V_1^2$ is also shown.  The DM annihilation
  rate is $\Gamma_a={1\over 2}C_AN^2$ where N is the number of WIMPs in the
  star.  Please note that the entry marked ``Sun'' refers to the
  present day Sun for comparison.  We have used the fiducial values
  $m_\chi=100\,{\rm GeV}$ and $\langle \sigma v\rangle_{\rm
    ann}=3\times 10^{-26}\,{\rm cm^3/s}$ as needed.}
\end{table*}

In Table 1, we have computed $C_A$ in Pop III stars of different
masses. From S. Woosley \cite{woosley}, we have obtained the
properties of zero metallicity stars when they are halfway through
hydrogen burning on the main sequence.  In the table we have used
our canonical values in Eq.(\ref{eq:anncrossection})
for the annihilation cross
section and mass ($m_\chi$ = 100 GeV); the results
easily be scaled to other values 
since $C_A \propto m_\chi^{1.5}\langle \sigma v
\rangle_{\rm ann}$.

\subsection{Capture Rate}
WIMP interactions with nuclei are of two kinds:
spin-independent, which scale as $A^2$ (where $A$ is the number of
nucleons in the nucleus), and spin-dependent, which require the
nucleon to have a spin.  Currently the experimental bounds on elastic
scattering are the weakest for the spin-dependent (SD) contribution
(to be precise we are considering only scattering off of protons since
the stars are comprised primarily of hydrogen). In figure 2
\cite{savage} we illustrate bounds on the SD component from direct
detection as well as from Super-Kamikande
\cite{savage,superk,xenon_si,zeplin,kims,cdms,DMLPG}.  The latter, which
are the most constraining, assume that a significant fraction of the
annihilation energy goes into neutrinos (as is likely for SUSY
particles); if the neutrino component is small then the SD cross
section could be several orders of magnitude higher.  As our fiducial
value, in this paper, we use the spin-dependent cross section
\begin{equation}
\label{eq:scattcrosssection}
\sigma_{c} = 10^{-39} {\rm cm}^2 
\end{equation}
which is consistent with all experimental bounds,
but we will always show the dependence of any result on the value
of $\sigma_c$.
The bound on the spin-independent (SI) scattering is much
tighter, $\sigma_{SI} \leq 10^{-42} {\rm cm}^2$
for $m_\chi = 100
$\,GeV \cite{cdms_si,xenon_si}.  The first stars are made only of
hydrogen or helium, so the $A^2$ enhancement for the spin-independent
contribution is not substantial.  In this paper we consider only the
spin-dependent contribution, though in principle for any specific
candidate WIMP one should self-consistently include both.  

The capture rate per unit volume at a distance $r$ from the center of
the star, for an observer at rest with respect to the WIMP
distribution (as should be a good approximation here), is
\cite{pressspergel,gould}
\begin{equation}
\label{eq:dcdv}
{d C \over dV}(r) = \left({6 \over \pi}\right)^{1/2} n(r) n_\chi(r) (
\sigma_c \bar v) {v(r)^2 \over {\bar v}^2} \left[ 1 - { 1-{\rm
exp}(-B^2) \over B^2} \right] ,
\end{equation}
where $n$ is the number density of nucleons (here, hydrogen), $n_\chi$
is the WIMP number density, $v(r)$ is the escape velocity of WIMPs
from the star at a given radius $r$, ${\bar v}^2 \equiv {3 k
  T_\chi \over M}$ is a ``velocity dispersion'' of WIMPs in the DM
halo, and
\begin{equation}
B^2 \equiv {3 \over 2} {v(r)^2 \over {\bar v}^2} {\mu \over \mu_-^2}
\end{equation}
where 
$
\mu = {m_\chi \over M_N}
$
is the ratio of WIMP to nucleon mass and
$
\mu_- = (\mu + 1)/2 .
$
For an observer moving with respect to the WIMPs, the quantity
in square brackets in Eq.(\ref{eq:dcdv})
becomes a more complicated function of $B$ and the
relative velocity as shown in Eq.(2.24) of \cite{gould}.

The capture rate for the entire star is then 
\begin{equation}
C = \int_0^{R_\star} 4 \pi r^2 dr {d C \over dV}(r) ,
\end{equation}
where $R_\star$ is the radius of the star.  To obtain a conservative
and fairly accurate estimate of the capture rate\footnote{We have subsequently obtained much more
  accurate values for $v(r)$ by treating the Pop III star as 
a polytrope with index $n=3$ (a good approximation), and confirmed that 
the conservative capture rate presented here differs (it is too low)
by only a factor of a few.}
, we may take
\begin{equation}
\label{eq:vescape}
v(r)^2 = v(R_\star)^2 = {2 G M_\star \over R_\star} \equiv v_{\rm esc}^2 
\end{equation}
for all $r$, assume the term in square brackets in Eq.(\ref{eq:dcdv})
is very close to 1 (justified below), and take a uniform dark matter
density.  In this case the integral simplifies to give
\begin{equation}
\label{eq:simplify}
C = \left({6 \over \pi} \right)^{1/2}\left({M_\star \over m_p}
f_H\right) (\sigma_c \bar v) \left({v_{\rm esc} \over {\bar
    v}}\right)^2 {\rho_\chi \over m_\chi} ,
\end{equation}
where $M_\star$ is the stellar mass, $m_p$ is the proton mass and
$f_H$ is the fraction of the star in hydrogen. (Note that hydrogen has
spin while helium generally does not.  We could, in principle,
consider the spin-independent contribution of scattering off of
hydrogen and helium in the stars; we have not done so because we
believe this contribution to be subdominant.  In any case the current
work is conservative in considering only the spin-dependent
scattering.)

To estimate $\bar v$ as per~\cite{binney}, we take the virial velocity
of the DM halo,
\begin{equation}
<{\bar v}^2 > = {|W| \over M_{\rm halo}},
\end{equation}
where 
\begin{equation}
W = - 4 \pi G \int \rho_{\rm halo} M_{\rm halo}(r) r dr
\end{equation}
and the typical DM halo containing a Pop III star has
$M_{\rm halo} = 10^5 - 10^6 M_\odot$. 

We use a Navarro, Frenk \& White (NFW) profile \cite{NFW} for the DM,
\begin{equation}
\label{eq:nfw}
\rho_{\rm halo} = {\rho_0 \over {r \over r_s} \left(1 + {r \over
    r_s}\right)^2 },
\end{equation}
where $r_s$ is the scale radius. The normalization $\rho_0$, known as
the central density, depends on the concentration parameter $C_{\rm
  vir}$ and on the redshift when the Halo virializes $Z_{\rm vir}$.
These parameters range from $C_{\rm vir} = (1-10)$ \cite{Diemand:2005vz}
and $Z_{\rm vir}=10-50$ \cite{Yoshida03,Reed:2005fb}.
With $r_s$ in the range (15-100) pc, we find $\bar v = (1-15)$\,km/s.
As our fiducial value, we will take
\begin{equation}
\bar v = 10\,{\rm km/s} .
\end{equation}

For $B \gg 1$, the term in square brackets in Eq.(\ref{eq:dcdv}) is
very close to 1, and this holds for all stellar and WIMP masses we are
considering. (For example, for a $1 M_\odot$ star, we have $v_{\rm esc}
= 618\,$km/s, which is much larger than $\bar v = 10\,$km/s; then for
$m_\chi = 100$ GeV we find $B \sim 100$.)  Thus we may ignore the term
in square brackets.  The bracketed term changes for a star moving
through the WIMP halo (rather than being stationary as we have
assumed), but the factor that replaces the term in square brackets is
$O(1)$. (For example, for a 1 $M_\odot$ star, with $m_\chi = 100$ GeV, and
with a velocity $\bar v$, we find that the factor is 0.66.)  We note that in
today's stars this factor is much more important than in the first
stars because $\bar v$ is much larger today (due to the fact that
today's galactic haloes are much larger, e.g. $10^{12}M_\odot$).

\begin{table*}
\begin{tabular}{lccccc}
  $M_\star(M_\odot)$ & $R_\star(R_\odot)$& $V_{\rm esc}(V_\odot)$ & $C (s^{-1})$&$\tau ({\rm yrs.})$ \\ \hline Sun &
  1&1&$4.9\times10^{34}$&$63$ \\ 10& $1.16$
  &$2.49$&$8.1\times10^{34}$&$152$ \\ 50&$4.76$	
  &$3.24$&$6.8\times10^{35}$&$190$ \\ 100& $7.04$
  &$3.77$&$1.9\times10^{36}$&$160$ \\ 250&$11.8$& $4.60$&
  $6.9\times10^{36}$&$126$\\
\end{tabular}
\caption{ \label{tab:PoPIII_Prop1} Stellar mass $(M_\star)$, radius
  $(R_\star)$, and surface escape velocity $(V_{\rm esc})$ in solar
  units, for metal free stars halfway through
  hydrogen burning.  We have also calculated the capture rate $C$
  using our fiducial values $\rho_\chi=10^9\,{\rm GeV/cm^3}$,
  $m_\chi=100\,{\rm GeV}$, and $\sigma_c=10^{-39}\,{\rm cm^2}$, and
  calculated $\tau$ using $C_A$ from Table 1. As in Table 1, the entry
  marked Sun refers to the present day Sun and not a zero metalicity
  star.  The capture rate for the Sun still uses the fiducial values;
  in the true present day Sun the true capture rate is much smaller
  ($C\approx10^{24}\,s^{-1}$), mostly due to the much lower DM
  densities in the solar neighborhood.}
\end{table*}

In Table 2, we evaluate the capture rate in Eq.(\ref{eq:simplify}),
again using properties (including stellar radius) 
of Pop III stars from \cite{woosley}.  In obtaining these
numbers, we have used $\rho_{\chi} = 10^9$ GeV/cm$^3$, $m_{\chi}$=100 GeV,
and $\sigma_c = 10^{-39} {\rm cm}^2$;
the result can easily be scaled to other values since
$C \propto \rho_{\chi} \sigma_c /m_{\chi}$.  
For general values we find that:
\begin{equation}
C =4.9\times10^{34} {\rm s}^{-1} \left({M_\star \over M_\odot}\right)
\left({v_{\rm esc} \over 618 {\rm km/s}} \right)^2 \left({\bar v \over
10 {\rm km/s}} \right)^{-1} \left({\rho_{\chi} \over 10^9 {\rm GeV /cm}}
\right) \left({m_{\chi} \over 100 {\rm GeV}} \right)^{-1} \left({
\sigma_{c} \over 10^{-39} {\rm cm}^2}\right) .
\end{equation}
Further, using Eq.(\ref{eq:vescape}) and noting that for the Pop III models
of \cite{woosley} it is roughly true that $R_\star \propto M_\star^{0.45}$,
we find that approximately
\begin{equation}
\label{eq:rough}
C \approx 4.9\times10^{34} {\rm s}^{-1} 
\left({M_\star \over M_\odot}\right) ^{1.55}
\left({\bar v \over
10 {\rm km/s}} \right)^{-1} \left({\rho_{\chi} \over 10^9 {\rm GeV /cm}}
\right) \left({m_{\chi} \over 100 {\rm GeV}} \right)^{-1} \left({
\sigma_{c} \over 10^{-39} {\rm cm}^2}\right).
\end{equation}

Table 2 also shows the equilibrium timescale given by Eq.(\ref{eq:equiltime})
using  the capture and annihilation rates determined above.
We can see that $\tau$ is extremely short, compared to the life time of a star.
Hence for most of the lifetime of the zero metallicity stars,
$t\gg\tau$ and we may use Eq.(\ref{eq:inequil}).

\subsection{Dark Matter Density}

To study the effects of dark matter on
the first stars, we need to know the density of the DM passing through
the stars to determine the capture rate.  Simulations have
unfortunately not (as yet) resolved this issue.  Below we will use a
variety of DM densities, since these numbers are unknown.
 
\subsubsection{Dark Matter Density before Capture is included}
First we need estimates of the DM density in the region where the star
forms, prior to including the effects of capture.  In a previous paper
\cite{sfg}, two of us (with P. Gondolo) used adiabatic contraction
\cite{Blumenthal:1985qy} to obtain estimates of the DM profile.  Prior
to this contraction, we assumed an overdense region of
$10^5M_\odot$--$10^6 M_\odot$ with a Navarro-Frenk-White (NFW) profile
\cite{NFW} for both DM and gas, where the gas contribution is 15\%.
(For comparison, we also used a Burkert profile \cite{burkert}, which
has a DM core. A Burkert profile has been shown to be a good fit for
the dynamics of today's galaxies\cite{gentile,salucci}) As the gas
collapses, we allowed the DM to respond to the changing baryonic
gravitational potential, where the gas density profiles were taken
from simulations of \cite{ABN,Gao06}). The final DM density profiles
were computed with adiabatic contraction [$M(r)r$ = constant].  After
contraction, we found a DM density at the outer edge of the baryonic
core of roughly $ \rho_\chi \simeq 5\,(n/{\rm cm}^{-3})^{0.81} {\rm
GeV/cm}^{-3} $ which scales as $\rho_\chi \propto r^{-1.9}$ outside
the core (see Fig.~1 in \cite{sfg}).  Our adiabatically contracted NFW
profiles match the DM profile obtained numerically in~\cite{ABN} (see
their Fig.~2).  They present their earliest (gas core density $n \sim
10^3 {\rm cm}^{-3}$) and latest ($n \sim 10^{13} {\rm cm}^{-3}$) DM
profiles, as far inward as $5 \times 10^{-3}$\,pc and $0.1$\,pc.  The
slope of these two curves is the same as ours.  If one extrapolates
them inward to smaller radii, one obtains the same DM densities as
with our adiabatic contraction approach.  The highest DM density found
by \cite{ABN} was $10^8$ Gev/cm$^{3}$.  Should the adiabatic contraction
continue all the way to the small stellar cores at $n\sim 10^{22} {\rm
cm}^{-3}$ (which we doubt), the DM density would be as high as
$10^{18} {\rm GeV/cm}^3$.  We note that \cite{sellwood} obtained DM
density profiles in galaxies and found that adiabatic contraction
produces densities that are too high by only a factor of 2 or 3, even
when radial orbits are included, or in the presence of bars, or in the
absence of spherical symmetry.  Below we will use a variety of DM
densities due to the uncertainties.  To give a sense of the numbers,
we compute that in a $10 M_\odot$ dark star, before capture is taken into
account, there are roughly $10^{16}$ baryons in the star for every
WIMP particle.

\subsubsection{Dark Matter Density Including Capture}
The amount of dark matter in the dark star obviously increases
significantly due to capture.  We find that (again for a $10 M_\odot$
star) there are $10^{12}$ baryons for every WIMP particle; i.e., the
fraction of WIMP particles has grown by a factor of $10^4$ due to
capture.  Of course the density of WIMPs is more centrally
concentrated, so that the ratio of WIMPs to baryons near the center of
the star, where the annihilation rate peaks, is higher than these
numbers.  Still, as we will show in the next section, it is remarkable
that particles which are $10^{-12}$ as numerous as the baryons can
provide the dominant heat source for the star.

\section{Luminosity due to WIMP annihilation:}
We may now compute the luminosity due to WIMP annihilation,
\begin{equation}
\label{eq:ldm}
L_{DM} = f \Gamma_A (2m_{\chi} ),
\end{equation}
where we take the energy per annihilation to be twice the WIMP mass;
two WIMPS annihilate per annihilation.  Here $f$ is the fraction of
annihilation energy that goes into the luminosity. Roughly 1/3 of the
annihilation energy is lost to neutrinos that stream right out of the
star, whereas the other 2/3 goes into electrons, positrons, and
photons that are trapped in the star and have their energy
thermalized. Hence we take \cite{sfg}
\begin{equation}
f \sim 2/3 .
\end{equation}
For $t\gg\tau$, which is quickly reached, Eq.(\ref{eq:ldm}) may be rewritten
using Eq.(\ref{eq:inequil}) as
\begin{equation}
L_{DM} = {f \over 2} C (2m_{\chi}) .
\end{equation}

\begin{table*}
\begin{tabular}{lccccc}
  $M_\star\,(M_\odot)$ & $L_\star\,({\rm Ergs/s})$& $L_{\rm DM}\,({\rm Ergs/s})$ & $\rho_\chi \,({\rm GeV/cm^3})$\ for which $L_\star=L_{DM}$ \\ \hline Sun &
     $3.9\times10^{33}$&$5.2\times 10^{33}$&$7.5\times10^{7}$ \\ 10& $4.2\times10^{37}$
  &$3.2\times10^{35}$&$1.3\times10^{10}$\\ 50&$2.0\times10^{39}$
  &$2.8\times10^{36}$&$7.5\times10^{10}$\\ 100& $6.45\times10^{39}$
  &$7.6\times10^{36}$&$8.3\times10^{10}$\\ 250&$2.31\times10^{40}$& $2.8\times10^{37}$&
  $8.5\times10^{10}$\\
\end{tabular}
\caption{ \label{tab:PoPIII_Prop3} Luminosity $L_\star$ due to fusion
  vs. luminosity $L_{DM}$ due to DM annihilation in zero metallicity
  stars (as defined in previous tables), using fiducial values for DM
  properties.  In the final column we vary $\rho_\chi$ to determine
  the value at which $L_{DM}$ exceeds $L_\star$. We stress that the DM
  densities listed here are those in the ambient medium (NFW plus
  adiabatic contraction) rather than the densities after capture; it
  is these ambient densities that determine the capture rate and thus
  the equilibrium luminosity.}
\end{table*}

The WIMP luminosity is given in Table III for a variety of stellar
masses together with the ordinary fusion-powered
stellar luminosity 
$L_\star$ provided by the models of \cite{woosley} for these stars.
Roughly, using Eq.(\ref{eq:rough}) we may write
\begin{equation}
\label{eq-ldm}
L_{DM} = 5.2\times 10^{33} {\rm erg/s} \left({M_\star \over M_\odot}
\right)^{1.55} \left({\rho_{\chi} \over 10^9 {\rm GeV/cm}^3} \right)
 \left({\sigma_{c} \over 10^{-39} {\rm cm}^2}\right)\left({\bar v \over
10 {\rm km/s}} \right)^{-1}.
\end{equation}

The WIMP luminosity depends linearly on the the WIMP density passing
through the stars.  We have also computed the WIMP energy density
$\rho_{\chi,\rm crit}$ that is required in order for the WIMP
annihilation energy to equal the ordinary stellar luminosity, which
will dramatically alter the properties of the first
stars~\cite{woosley2}. We stress that the DM densities listed here are
those in the ambient medium (NFW plus adiabatic contraction) rather
than the densities after capture; it is this ambient density that
determines the capture rate, and also therefore the the annihilation
rate (in equilibrium), and consequently the luminosity.  For any WIMP
densities higher than this value, the star's luminosity is dominated
by annihilation energy (rather than by ordinary fusion).  We note that
the luminosity due to fusion in the zero-metallicity stars
\cite{woosley} scales as $L_{\star} \propto M^2$ (see figure 1)
whereas $L_{DM} \propto M^{1.55}$, so that DM heating is relatively
more important in lower-mass stars.  The DM luminosity is dominant for
the values of $\rho_{\chi,\rm crit}$ given in the last column of Table
III, and is dominant for all relevant stellar masses for $\rho_{\chi}
\gtrsim 9 \times 10^{11}$GeV/cm$^3$; thus for dark matter densities in
excess of this value, it will be the DM heating that determines
stellar properties, and the star is a dark star.  
The two luminosities are plotted in Figure 1
where they can be compared with one another.  If the first stars are
observed (e.g., by James Webb Space Telescope) to have the properties
predicted of fusion-driven stars, then one could use these results to
constrain WIMP properties to make sure that DM annihilation remains
subdominant.

A potentially important consideration is the time for which a dark star
will be able to burn DM at the calculated rate, given that it will
annihilate its captured store of DM in a time of order $\tau$ as given
by Eq.(\ref{eq:equiltime}). The maximal burning time is determined by
the total mass of DM in the region of phase space that intersects the
star.  This mass could be small if the star were fixed exactly at the
central cusp of the DM halo.  However, we expect that given the
complexities of the collapse process, the star will have some nonzero
velocity and thus `wander' through a region of some radius
significantly larger than that of the star.  This makes much more DM
mass available for burning, at the expense of somewhat lower average
DM density.  (It should then be noted that the fiducial DM density
numbers we assume are not those expected in the innermost core but
over some significantly larger region.)  Estimation of the expected
degree of wandering, the detailed DM density profile, and therefore
the maximal burning time of the dark stars
will be left for future and more detailed study.

\begin{figure}[b]
\centerline{\includegraphics[width=0.5\textwidth]{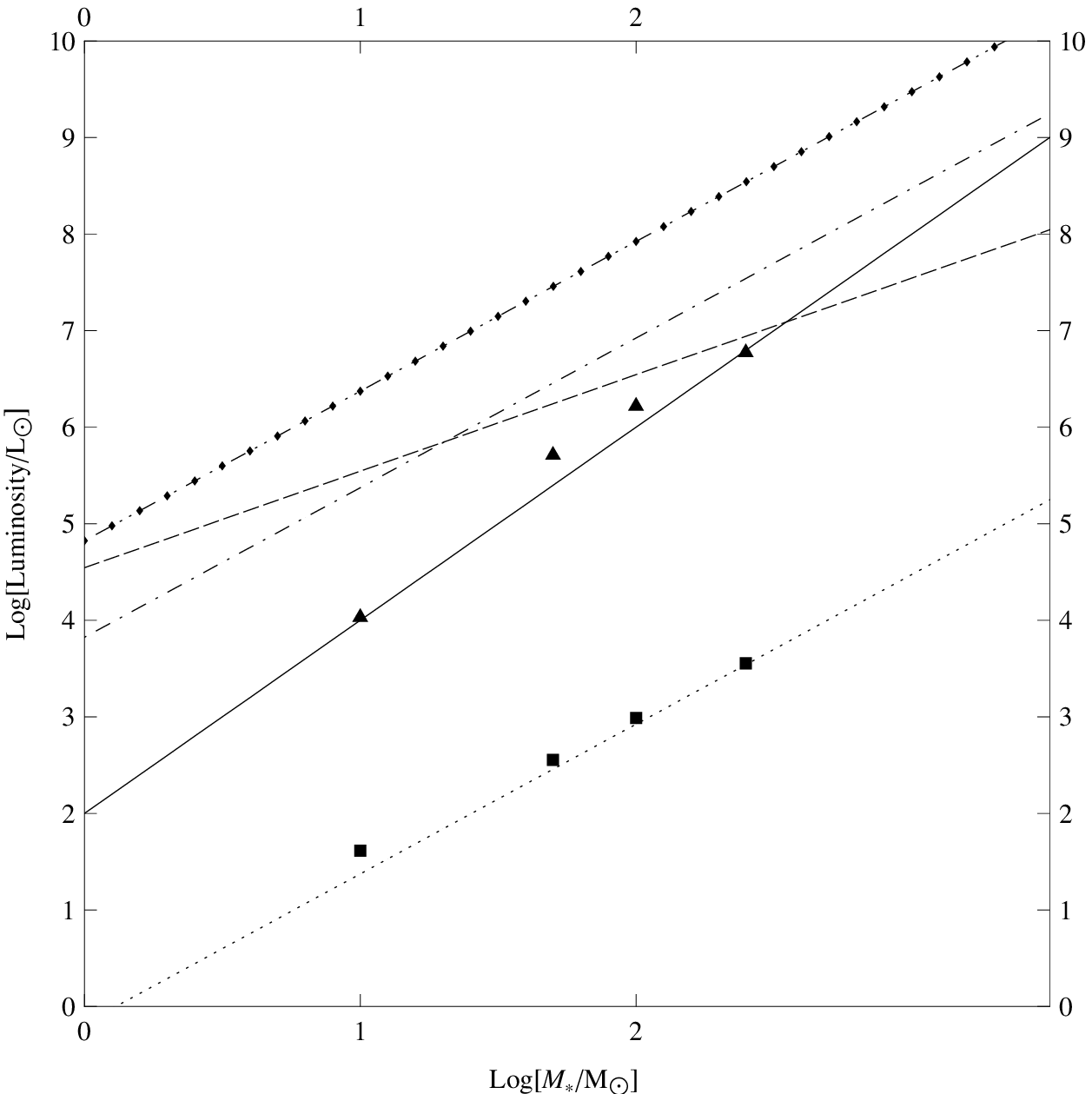}}
\caption{Log-Log plot of stellar luminosity vs.  stellar mass (in
solar units). The dashed line shows Eddington luminosity for Thompson
scattering.  Triangles show the data points for luminosity of zero
metallically stars in Table 3; the solid line is a fit to these
points.  The remaining lines indicate the DM annihilation luminosity
for different DM density assumptions.  We stress that the DM densities listed
here are those in the ambient medium (NFW plus adiabatic contraction)
rather than the densities after capture; it is these
densities that determine the capture rate and the luminosity.
The dotted line is our fiducial
example with $\rho_\chi=10^9 \,{\rm GeV/cm^3}$ and
$\sigma_c=10^{-39}\,{\rm cm^2}$ (squares are from the third column of
Table 3.) If we keep the cross section fixed, the middle dot-dashed
line corresponds to a DM density of $10^{12}\,{\rm GeV/cm^3}$, and the
top diamond line would have a DM density of $10^{14}\,{\rm GeV/cm^3}$;
in this final case, the DM luminosity would dominate over the
Eddington luminosity for a star with a mass of $\leq1\,M_\odot$.
\vspace{-0.5\baselineskip} }
\end{figure}

\section{Eddington luminosity}

We as yet know little about the effect that significant dark matter
heating would have on the structure of zero-metallicity stars, but we
can nonetheless place an approximate upper bound on the stars' mass if
we assume that they must be sub-Eddington. In the current paper, we
simply take the ZAMS stars, add the DM luminosity, and ask whether or
not the resulting dark stars are self-consistent.  In other words, is
it possible for dark stars of this mass to exist? We compute the
Eddington luminosity $L_{\rm Edd}$ for these objects, and ascertain
whether or not the DM luminosity is in excess of this value.  If
$L_{\rm DM} > L_{\rm Edd}$, then the star of this mass cannot exist:
the pressure from the star would be so large as to blow off some of
the mass.  In reality one should do a different problem:
one should really follow the protostars and compute their structure as they
accrete mass on their way to becoming ZAMSs.
Such a calculation would essentially
compare accretion luminosity (the value of which would depend on the
nature of the accretion, e.g. spherical or disk accretion) to the
Eddington luminosity; here we are comparing dark star luminosity to
the Eddington luminosity.

The Eddington luminosity is
defined (e.g.~\cite{hansen}) to be
\begin{equation}
\label{eq-ledd}
L_{\rm Edd}=\frac{4\pi cG M_\star}{\kappa_p},
\end{equation}
where $G$ is Newton's Constant, $c$ is the speed of light, $M_\star$
is the mass of the star, and $\kappa_p$ is the opacity of stellar
atmosphere.  Since the first stars' stellar atmospheres are hot and
nearly metal-free, the opacity is dominated by Thompson scattering, so
we take
\begin{equation}
L_{\rm Edd} = 1.4 \times 10^{38} {\rm erg/s}\, (M_\star/M_\odot) =
3.5\times10^4 (M_\star/M_\odot) L_\odot.
\end{equation}

In Figure 1, we have plotted three luminosities as a function of
stellar mass for zero metallicity stars: the luminosity due to fusion,
the luminosity due to WIMP annihilation (for a variety of WIMP
densities in the star), and the Eddington luminosity.  Since the
Eddington luminosity scales as $L \propto M_\star$ whereas the DM
luminosity scales as $L_{\rm DM} \propto M_\star^{1.55}$, for a large
enough dark matter density the two curves will cross for some stellar
mass.  The lightest stellar mass for which $L_{\rm DM} > L_{\rm Edd}$
then constitutes an approximate upper mass limit to the first stars,
because the star of that mass is unable to accrete any further due to
the radiation pressure from the WIMP annihilation.  If one assumes
that accretion efficiently drives up the mass of any Pop III star,
then the mass of the first star will be uniquely determined by the
properties of the DM. Using Eqs.(\ref{eq-ldm}) and (\ref{eq-ledd}), we
get:
\begin{equation}
M_\star^{\rm max} = 1.1\times 10^8\,M_\odot\left({\rho_\chi\over
10^9\,{\rm GeV/cm^3}}\right)^{-1.8}\left({\sigma_{c} \over 10^{-39}
{\rm cm}^2}\right)^{-1.8}.
\end{equation} 

For example, we find that for a dark matter density of $\rho_{\chi} =
1.1\times10^{13}$ GeV/cm$^3$, the first stars cannot be more massive
than $1 M_\odot$.  Figure 2 illustrates $M_\star^{\rm max}$ as a
function of WIMP energy density for several values of the scattering
cross section.

\begin{figure}[b]
\centerline{\includegraphics[width=0.5\textwidth]{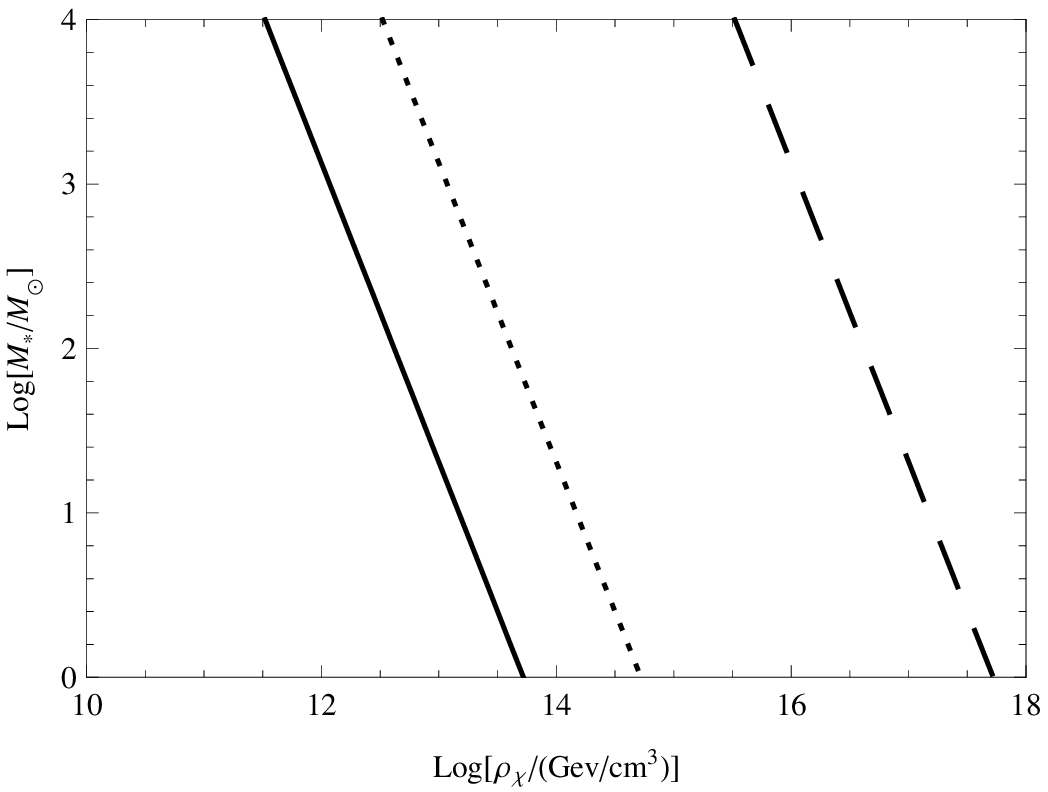}}
\caption{Maximum stellar mass due to the
Eddington limit as a function of $\rho_\chi$ for a fixed cross
section.  The different lines correspond to different spin-dependent
cross sections: $\sigma_c=10^{-39}\,{\rm cm^2}$ (solid line),
$\sigma_c=10^{-41}\,{\rm cm^2}$ (dotted line), and
$\sigma_c=10^{-43}\,{\rm cm^2}$ (dashed line).  }
\end{figure}

In principle, these arguments can be turned around to place a bound on the
scattering cross section.  Figure 3 is a plot of WIMP scattering cross
section vs. WIMP mass.  Experimental bounds are shown.  The horizontal
lines indicate the values of $\sigma_c$ that correspond to different
values of $\rho_{\chi}$ at which a 1 $M_\odot$ star is Eddington
limited by DM annihilation.  (The scaling to other stellar masses is
straightforward since $L_{DM} \propto M_\star^{1.55}$.)  Once the mass
of the first stars is determined, then one could rule out any
combination of $\sigma_c$ and $\rho_{\chi}$ for a given WIMP mass that
would preclude stars of such a mass from forming.  As an extreme
example, if one were to believe adiabatic contraction all the way to
the limit where the protostellar core has gas density $n \sim
10^{22}\, {\rm cm}^{-3}$, the WIMP density would reach $\rho_{\chi} =
10^{18}\, {\rm GeV/cm}^3$.  In this case the Eddington limit would be
reached for masses $\ll 1\, M_{\odot}$ for our fiducial scattering
cross section.  If, instead, the first stars are observed to form with
masses larger than one solar mass, and such an enormous WIMP density
were found to be sensible, one could place a bound of $\sigma_c <
3\times10^{-44}\, {\rm cm}^2$ on WIMPs for almost any mass (the
equilibrium DM luminosity is independent of $m_\chi$), which would be
the tightest known bound on the spin dependent scattering cross
section by several orders of magnitude; see figure 3. In this extreme
case we could also put interesting limits on spin independent
scattering; see figure 4.  

Obtaining bounds on the WIMP parameters in this way depends on
detailed understanding of the ambient WIMP density within which the
dark star resides.  Clearly this is a very difficult problem which
will not be solved in the near future.  Indeed, it is likely that the
$10^6 M_\odot$ haloes containing the dark stars will merge with other
haloes and that the stars will not remain forever in regions of high
DM density.  However, simulations of structure formation are becoming
ever better and one may hope someday to address this question. 

\begin{figure}[b]
\centerline{\includegraphics[width=0.5\textwidth]{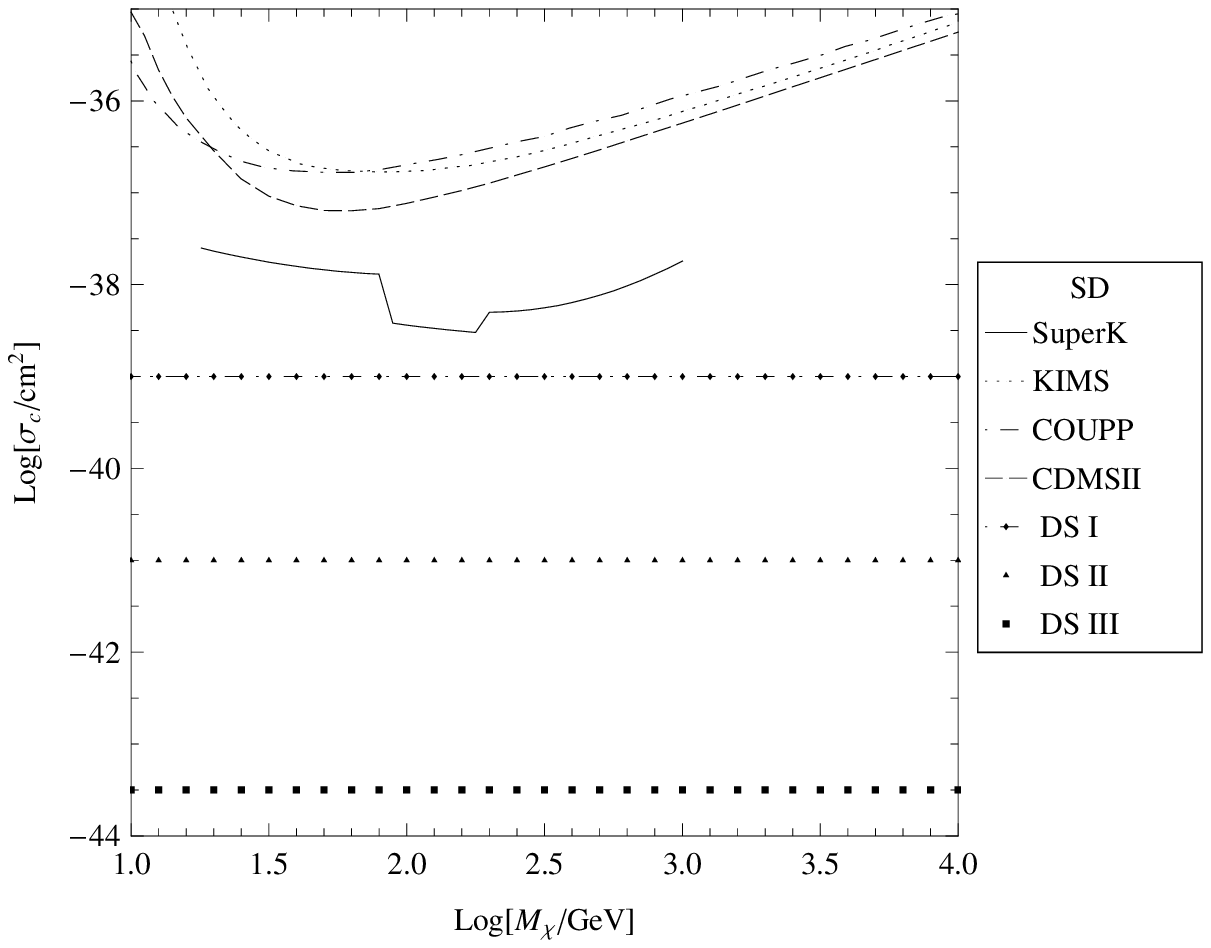}}
\caption{Bounds on spin-dependent scattering cross section as a function
of WIMP mass from experiments, as well as potential bounds from the
effect of WIMP annihilation on the first stars. Displayed limits
include various direct and indirect detection experiments as labeled;
the tightest present bounds on spin-dependent cross-sections are from
Super-Kamiokande (labeled SuperK), at around $10^{-38}\,{\rm cm^2}$
\cite{savage,superk,kims,cdms,DMLPG}.  The three horizonal lines
are bounds that would result from the discovery of Pop III stars of $1
M_\odot$ for a variety of DM densities inside these stars (due to the
fact that DM annihilation pressure would otherwise prevent their
formation).  The lines are labeled as follows: DS I 
corresponds to ambient density $\rho_\chi=1.1\times10^{13}\,{\rm
GeV/cm^3}$  and would lead to the bound
$\sigma_c\leq10^{-39}\,{\rm cm^2}$; DS II corresponds to
$\rho_\chi=3\times10^{15}\,{\rm GeV/cm^3}$ and would lead to
$\sigma_c\leq10^{-41}\,{\rm cm^2}$; and DS III corresponds to
$\rho_\chi=10^{18}\,{\rm GeV/cm^3}$ and would lead to 
$\sigma_c<3\times10^{-44}\,{\rm cm^2}$.
\vspace{-0.5\baselineskip}
}
\end{figure}

\begin{figure}[b]
\centerline{\includegraphics[width=0.5\textwidth]{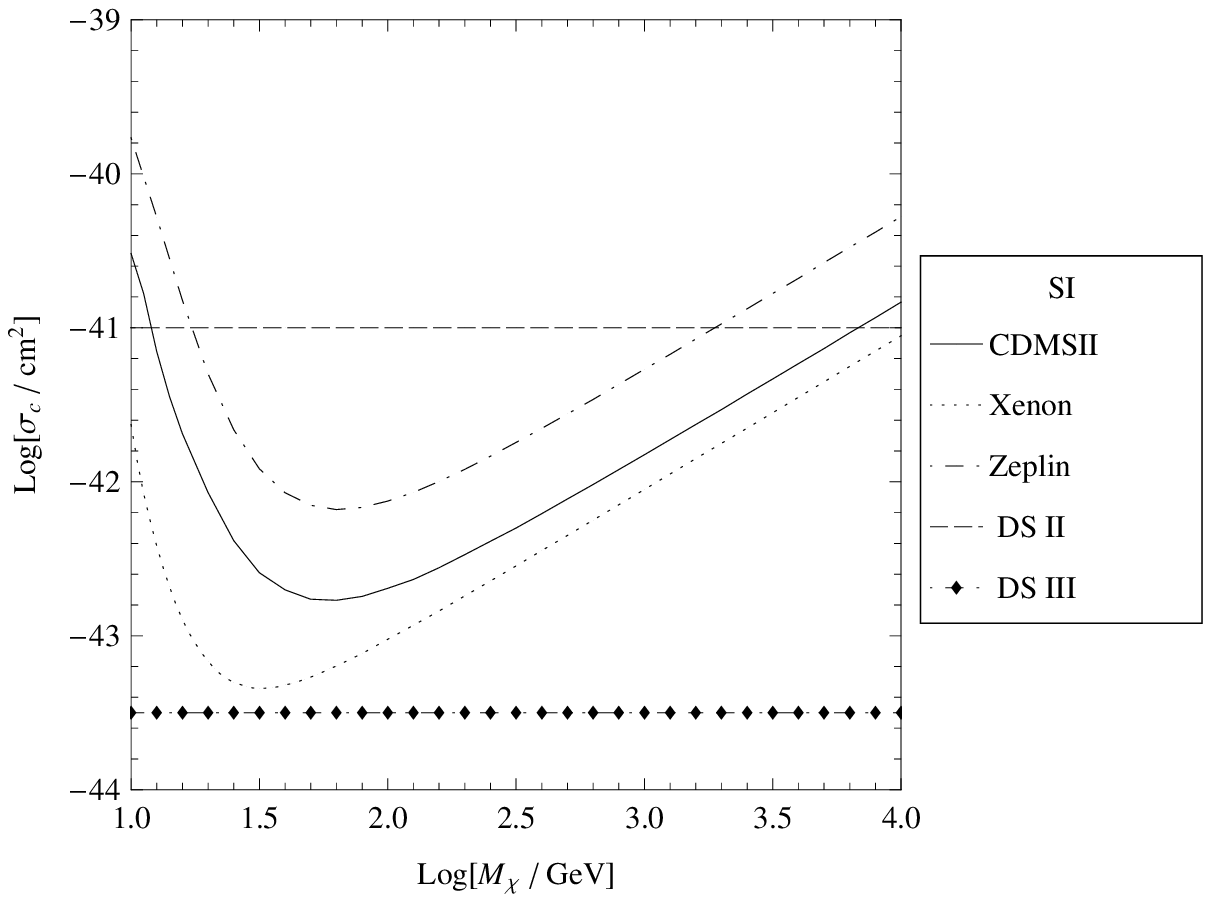}}
\caption{Same as previous figure but for WIMPs with spin-independent 
interactions
\cite{cdms_si,xenon_si, zeplin_si,DMLPG}.
\vspace{-0.5\baselineskip}
}
\end{figure}

\section{Conclusion}

In summary, Pop. III stars are expected to reside at the core of a
$10^5-10^6\,M_\odot$ dark matter halo.  Previously two of us
\cite{sfg} discussed the importance of dark matter annihilation in the
first stars, proposing the existence of ``dark stars'' powered by DM
heating.  Even once the DM initially collapsing with the baryons into
the star runs out as a source of fuel, it can be replenished by
capture of more DM from the ambient medium (as the DM passes through
the dark star).  In this paper, we have estimated the rates of WIMP
capture and self-annihilation in such stars. We have shown that when
these rates are in equilibrium, the accompanying heating would provide
an energy source that can rival nuclear fusion and prolong the dark
star phase, if the core dark matter density is sufficiently high.
Such densities seem plausible based on analytic models of
adiabatically contracted halos \cite{sfg}, suggesting that DM heating
may radically affect the structure of the first stars.

The two key uncertainties in this work are: (i) the scattering cross
section must be at (or near) the experimental limit, and (ii) the
ambient DM density must be high enough for capture to take place. The
second criterion is likely to be true for a while after the star is
created, but it is not clear for how long.  Once the halo containing
the Dark Star merges with other objects, it is not clear how long the
central DM in the halo remains undisturbed, and it is not clear how
long the Dark Star remains at this central point.  In principle the
capture could continue indefinitely so that a dark star could still
exist today, but this is very unclear.

For high enough DM density, DM heating will lower the Eddington
stellar mass limit to provide an upper mass cutoff for Pop. III;
because DM heating might also affect the formation properties of
Pop. III stars, the gross properties of these objects may well be
determined by the particle properties of dark matter.  And conversely,
inferred properties of Pop. III stars (or even future direct
observations) might be used to strongly constrain DM masses and
interaction cross sections.

We thank S. Woosley for sharing with us his models of zero metallicity
stars. We also thank C. Church, J. Primack, S. Profumo and C. Savage
for useful discussions.  We acknowledge support from: the DOE and MCTP
via the Univ.\ of Michigan, the Perimeter Institute (K.F.), NSF grant
AST-0507117 and GAANN (D.S., A.A.). K.F. acknowledges the hospitality
of the Physics Dept. at the Univ. of Utah.

\end{document}